\documentclass[12pt]{article}
\usepackage{epsf}
\def\beq{\begin{equation}}
\def\eeq{\end{equation}}
\def\ap#1#2#3 {Ann. Phys. (NY) {\bf#1} (19#2) #3}
\def\err#1#2#3 {{\it Erratum} {\bf#1} (19#2) #3}
\def\ib#1#2#3 {{\it ibid.} {\bf#1} (19#2) #3}
\def\ijmp#1#2#3 {Int. J. Mod. Phys. {\bf#1} (19#2) #3}
\def\jetp#1#2#3 {JETP Lett. {\bf#1} (19#2) #3}
\def\mpl#1#2#3 {Mod. Phys. Lett. {\bf#1} (19#2) #3}
\def\np#1#2#3 {Nucl. Phys. {\bf#1} (19#2) #3}
\def\pl#1#2#3 {Phys. Lett. {\bf#1} (19#2) #3}
\def\prep#1#2#3 {Phys. Rep. {\bf#1} (19#2) #3}
\def\prev#1#2#3 {Phys. Rev. {\bf#1} (19#2) #3}
\def\prl#1#2#3 {Phys. Rev. Lett. {\bf#1} (19#2) #3}
\def\sjnp#1#2#3 {Sov. J. Nucl. Phys. {\bf#1} (19#2) #3}
\def\spj#1#2#3 {Sov. Phys. JETP {\bf#1} (19#2) #3}
\def\spu#1#2#3 {Sov. Phys. Usp. {\bf#1} (19#2) #3}
\def\zp#1#2#3 {Zeit. Phys. {\bf#1} (19#2) #3}

\def\dsb{{\bar D}^{*0}}
\def\x{X(3872)}
\def\lsim{\raisebox{-0.6ex}{$\stackrel{\textstyle <}{\sim}$}}

\begin{document}
\begin{titlepage}
\begin{center}
{\Large \bf William I. Fine Theoretical Physics Institute \\
University of Minnesota \\}  \end{center}
\vspace{0.2in}
\begin{flushright}
TPI-MINN-03/27-T \\
UMN-TH-2216-03 \\
September 2003 \\
\end{flushright}
\vspace{0.3in}
\begin{center}
{\Large \bf  Interference and binding effects in decays of possible
molecular component of $X(3872)$
\\}
\vspace{0.2in}
{\bf M.B. Voloshin  \\ }
William I. Fine Theoretical Physics Institute, University of
Minnesota,\\ Minneapolis, MN 55455 \\
and \\
Institute of Theoretical and Experimental Physics, Moscow, 117259
\\[0.2in]
\end{center}

\begin{abstract}
It is pointed out that the internal structure of the narrow resonance
$X(3872)$ at the $D^0 {\bar D}^{*0}$ threshold can be studied in some
detail by measuring the rate and the spectra in the decays $X(3872) \to
D^0 {\bar D}^0 \pi^0$ and $X(3872) \to D^0 {\bar D}^0 \gamma$. In
particular, if this resonance contains a dominant `molecular' component
$D {\bar D^*} \pm {\bar D} D^*$, this component can be revealed and
studied by a distinct pattern of interference between the underlying
decays of $D^{*0}$ and ${\bar D}^{*0}$ whose coherence is ensured by
fixed (but yet unknown) $C$ parity of the $X(3872)$.
\end{abstract}

\end{titlepage}

The recently observed\cite{belle} by the Belle Collaboration narrow
resonance $X(3872)$ at $3872.0 \pm 0.6 \pm 0.5 \, MeV$ (and confirmed by
CDF\cite{cdf} at $3871.4 \pm 0.7 \pm 0.4 \, MeV$) decaying into $\pi^+
\pi^- J/\psi$ is within $0.2 \pm 0.7 \, MeV$ from the $D^0 {\bar
D}^{*0}$
threshold. The extreme proximity of the resonance to the threshold
naturally invites the suggestion\cite{tornquist, cp} that its wave
function may have a large component with a pair of neutral pseudoscalar
and vector $D\, ({\bar D})$ (anti)mesons. The spatial separation of the
mesons in this component is sufficiently large for the mesons to retain
their individual structure. Such configuration would clearly realize the
long-standing conjecture\cite{vo,dgg} of existence of ``molecular
charmonium" i.e. of resonances, which essentially are loosely bound
states of charmed hadrons.

Clearly, the approximately $1 \, MeV$ or less scale for the energy gap
$w$ between the resonance and the threshold is quite likely to result in
a completely different weight in the wave function of the $X(3872)$ of
the pairs of neutral and charged $D$ mesons, since the threshold for the
charged $D^+ {\bar D}^{*-}$ pairs is another $8.1 \pm 0.5 \, MeV$
higher,
which is `far' in the scale of $w$. Thus the isospin is likely to be
strongly violated in the $\x$ resonance, which in the case if $\x$ is
even under $C$ parity, would allow the observed decay $\x \to \pi^+
\pi^- J/\psi$ to be in fact occurring due to the decay $\x \to \rho^0
J\psi\,$\cite{tornquist,cp}, in agreement with the very strong peaking
at the maximal value of the spectrum of the invariant mass of the two
pions\cite{belle}.  One can trivially notice that this conjecture can be
readily tested by a search for decay involving neutral pions: $\x \to
\pi^0 \pi^0 J/\psi$. If indeed the pions emerge from the $\rho^0$
resonance, the process with neutral pions should be absent. The decay
with neutral pions is also forbidden, and the observed decay $\x \to
\pi^+ \pi^- J/\psi$ is manifestly due to an $I=1$ component of $\x$, in
the general case of $C(X)=+1$,  even if the $\rho^0$ dominance is not
confirmed. Alternatively, if the  discussed resonance is a $C=-1$ state,
the decay $\x \to \pi^0 \pi^0 J/\psi$ is allowed with the dipion being
in the $I=0$ isospin state, and the relation $\Gamma(X \to \pi^+ \pi^-
J/\psi)= 2 \, \Gamma(X \to \pi^0 \pi^0 J/\psi)$ should hold to a good
accuracy. (Any significant presence of an $I=2$ state of the dipion
would obviously be totally exotic.)

Naturally, further study of the properties of $\x$ will likely involve
other possible decays of this resonance, including the decays related to
the underlying transitions $D^{*0} \to D^0 \pi^0$, $D^{*0} \to D^0
\gamma$, and the corresponding transitions between the
anti-mesons\cite{tornquist,cp}.
The main purpose of the present paper is to point out that at the
characteristic momenta of the mesons in the wave function of the
molecular $D^0 \dsb$ component of the $\x$ resonance the parameters of
these decays should likely be measurably different from those of an
incoherent sum of decays of free $D^{*0}$ and $\dsb$ mesons.
Rather the rates and the spectra of the decays $X(3872) \to D^0 {\bar
D}^0 \pi^0$ and $X(3872) \to D^0 {\bar D}^0 \gamma$ should exhibit
binding effects and a significant interference between the underlying
decays of vector mesons and anti-mesons. Thus an experimental study of
these decays may reveal rather fine details of the structure of the $\x$
resonance. In other words, the Dalitz plots of these decays would
provide a ``CAT scan" of the actual wave function of the mesons inside
$\x$.

It can be noted, that the spatial (momentum) dependence of the main part
of the wave function of the mesons can be described, in a way, similar
to that used for deuteron\cite{tornquist}, and the essential unknown
parameter for this part is the overall normalization, which represents
the weight of the molecular component in the wave function of the $\x$
resonance. Indeed,
at the gap energy $w \, \lsim \, 1 \, MeV$ the dynamics of the $D^0
\dsb$ (${\bar D}^0 D^{*0}$) meson pair is determined by momenta of order
$\kappa = \sqrt{2 \mu |\epsilon|} \, \lsim \, 45 \, MeV$, where $\mu
\approx 966 \, MeV$ is the reduced mass of the system made of
pseudoscalar and vector neutral $D$ mesons. Thus the characteristic
distances $\kappa^{-1}$ are far beyond the range of the strong
interaction, and the wave function at those characteristic distances is
in fact given by the Schr\"odinger equation for free motion. On the
other hand the value of $\kappa$ may well be comparable with the
momentum $p$ of the pion emitted in the decay $X(3872) \to D^0 {\bar
D}^0 \pi^0$ ($p_0 = 43 \, MeV$ for a decay of a free $D^{*0}$ meson),
which would give rise to large binding and interference effects in the
decay. In the case of radiative decay  $X(3872) \to D^0 {\bar D}^0
\gamma$, the representative value of the photon momentum $k$ is that in
a free $D^{*0}$ decay: $k_0=137 \, MeV$. Although this value looks large
as compared to $\kappa$, it will be shown that the relative magnitude of
the interference effect is determined by the expression $ (2 \kappa/k)
\, \arctan (k/2\kappa)$ and is significant for this decay as well.

In the following discussion it is assumed for definiteness that $\x$ is
below the $D^0 \dsb$ threshold, so that $w = m+M-M(X)$ is a positive
quantity, where $m=M(D^0)\approx 1864.5 \, MeV$ and $M=M(D^{*0}) \approx
2006.7 \, MeV$. A generalization to the case where $\x$ is just above
the threshold can be done by analytical continuation. Also for
definiteness it is assumed here that the mesons inside the $\x$ are in
the $S$ wave, which quite plausibly is the actual situation. This
obviously corresponds to $J^{PC}(X)$ equal to either $1^{++}$, or
$1^{+-}$. If it further turns out that the quantum numbers of $\x$ are
different, the orbital motion of the mesons can readily be accounted for
by a straightforward modification of the formulas presented below. Thus
the wave function of the relative motion of the mesons is considered
here as given by the standard $S$ wave expression
\beq
\psi({\vec r})=\xi \, \sqrt{\kappa \over 2 \pi} \, {e^{-\kappa r} \over
r}~,
\label{swave}
\eeq
where $\xi^2$ is the overall weight of the considered here molecular
component in the $\x$ resonance. For a purely molecular system
$\xi^2=1$, while realistically one would expect $\xi^2 < 1$ thus
allowing for some admixture in the wave function of $\x$ of other states
(e.g. $c {\bar c}$, $D^+ D^{*-}$, etc.).

The amplitude for the decay $D^{*0} \to D^0 \pi^0$ can be written as
\beq
A_{D^*D\pi}=g \, ({\vec \epsilon} \cdot {\vec p})~,
\label{addpi}
\eeq
where ${\vec \epsilon}$ is the polarization amplitude of the $D^*$ and
${\vec p}$ is the momentum of the pion. The coupling constant $g$ is
related to the width $\Gamma_\pi \equiv \Gamma(D^{*0} \to D^0 \pi^0)$
as\footnote{The nonrelativistic normalization is used here for the wave
functions of the $D$ mesons, but not for the pion.}  $\Gamma_\pi=|g^2|
p_0^3/6\pi$. The rate $\Gamma_\pi$ can be estimated from the isotopic
symmetry and the known\cite{pdg} total width of $D^{*+}$ ($96 \pm 22 \,
KeV$) and the branching ratio $B(D^{*+} \to D^+ \pi^0)=(30.7 \pm 0.5) \%
$, and also taking into account the slight difference in the kinematics:
$\Gamma_\pi = 43 \pm 10 \, KeV$. In terms of the coupling $g$ this leads
to a quite reasonable estimate $|g^{-1}| = 315 \pm 36 \, MeV$.

For a system of a vector and a pseudoscalar mesons, with a definite $C$
parity $\eta$ the amplitude of decay into $D^0 {\bar D}^0 \pi^0$ is
contributed by both the decay $D^{*0} \to D^0 \pi^0$ and its
charge-conjugate $\dsb \to {\bar D}^0 \pi^0$. Taking into account that
$C(\pi^0)=+1$, and performing the standard transition to the
center-of-mass coordinate ${\vec R}$ and the relative coordinate ${\vec
r}$ the amplitude of the decay $\x \to D^0 {\bar D}^0 \pi^0$ can be
written as
\begin{eqnarray}
&&\langle D^0({\vec q_1}) {\bar D}^0({\vec q_2}) \pi^0({\vec p})| H_{D^*
D \pi}| X ({\vec \epsilon}, {\vec P=0}) \rangle = \nonumber \\
&&(2 \pi)^3 \delta^{(3)} ({\vec q_1}+{\vec q_2}+{\vec p}) \, g \, ({\vec
\epsilon} \cdot {\vec p}) \, \left [ \phi({\vec q_2}) + \eta \,
\phi({\vec
q_1}) \right ]~,
\label{axddpi}
\end{eqnarray}
where ${\vec q_1}$ (${\vec q_2}$) is the momentum of the final $D$
(${\bar D}$) meson, ${\vec p}$ is the pion momentum, and ${\vec P}$ is
the momentum of the initial $\x$ resonance, which is set to zero,
corresponding to consideration in the rest frame of the $\x$. (In
eq.(\ref{axddpi}) a use is made of the momentum conservation relation in
this specific frame: ${\vec q_1}+{\vec q_2}+{\vec p}=0$, which somewhat
simplifies the formula and the subsequent discussion.) Finally,
$\phi({\vec q})$ is the wave function of the relative motion in the
momentum representation:
\beq
\phi({\vec q}) = \int \psi^*({\vec r}) \, e^{i {\vec q} \cdot {\vec r}}
\,  d^3 r~.
\label{phiq}
\eeq
The `free' wave function in eq.(\ref{swave}) in the momentum space reads
as
\beq
\phi({\vec q})= \xi \, {\sqrt{8\, \pi \, \kappa} \over q^2+\kappa^2}~.
\label{msw}
\eeq

It should be noted, that in the expression in eq.(\ref{axddpi}) it is
assumed that the final $D$ and ${\bar D}$ mesons move as free particles,
i.e. that the wave function of each is a plane wave $\exp(i {\vec q}
\cdot {\vec r})$. Such assumption looks quite reasonable, since the
final $D$ mesons are produced at large distances of order $\kappa^{-1}$
from each other, i.e. beyond the range of strong interaction. This
behavior would be invalid if there were a resonance or a bound state of
the pseudoscalar $D$ mesons very close to their threshold, similar to
the $\x$ state at the threshold of $D^0 \dsb$. Existence of such
resonance would certainly be a new phenomenon by itself, and would
require a separate consideration. Here it is assumed that no singularity
exists in the spectrum of $D {\bar D}$ pairs within at least few MeV
near their threshold.

Using eq.(\ref{axddpi}) the expression for the decay rate can be written
in terms of $\phi({\vec q})$ in the textbook form:
\begin{eqnarray}
&&d \Gamma(X \to D^0 {\bar D}^0 \pi^0)= \nonumber \\
&& |g^2| \, {p^2 \over 3 \, (2 \pi)^5} \, \left | \phi({\vec q_2}) +
\eta \, \phi({\vec q_1}) \right | ^2 \, \delta \left (\Delta -
w - E_\pi - {q_1^2 \over 2 m }- {q_2^2 \over 2 m}  \right ) \,
\delta^{(3)}  \left ( {\vec q_1}+{\vec q_2}+{\vec p} \right ) \, d^3
q_1 \, d^3 q_2 {d^3 p \over 2 E_\pi } \nonumber \\
&&=|g^2| \, {p^2 \over 96 \pi^3} \, \left | \phi({\vec q_2}) + \eta \,
\phi({\vec q_1}) \right |^2 \, dq_1^2 \, dq_2^2~,
\label{gam}
\end{eqnarray}
where $\Delta=M-m =142.12 \pm 0.07 \, MeV$ is the difference between the
masses of the vector and pseudoscalar neutral $D$ mesons, and $E_\pi =
\sqrt{p^2+m_\pi^2}$ is the energy of the pion. The intermediate
expression with un-integrated delta-functions is convenient for
discussing the limiting case of loose binding, $\kappa \to 0$, while the
final one is the standard Dalitz type and thus is convenient for
discussing the decay parameters in terms of the Dalitz plot. (Clearly,
in the latter expression the value of $p^2$ is uniquely determined
through the conservation laws by the values of $q_1$ and $q_2$.)

In the limit of no binding ($\kappa \to 0$) the momentum space wave
function can be replaced as $|\phi({\vec q})|^2 \to \xi^2 (2\pi)^3
\delta^{(3)} ({\vec q})$, and the intermediate expression splits into
two noninterfering terms, corresponding to independent `free' decays
$D^{*0} \to D^0 \pi^0$ $({\vec q_2} =0)$, and $\dsb \to {\bar D}^0
\pi^0$ $({\vec q_1} =0)$, thus recovering the naive expression for the
total width $\Gamma = 2 \, \xi^2 \, \Gamma_\pi$ (and the trivial
kinematics).
However at a value of $\kappa$ comparable with $p_0$ the spread of the
momentum space wave function and the interference effects are essential.

In order to present the results of calculation of from eq.(\ref{gam}),
we write the total rate of the discussed decay in the form
\beq
\Gamma(X \to D^0 {\bar D}^0 \pi^0)=2 \xi^2 \, \Gamma_\pi \, \left [ A(w)
+ \eta \, B(w) \right ]~,
\label{gamt}
\eeq
where $A(w)$ describes the incoherent contribution of the decays of
individual $D^{*0}$ and $\dsb$, and $B(w)$ describes the effect of the
interference between these two processes.
The result of a numerical calculation of the terms $A$ and $B$ with the
wave function from eq.(\ref{msw}) is shown in Fig.1. It is seen from the
plot, that the discussed effects reach quite sizeable magnitude starting
already from small values of the binding energy $w \sim 0.1 \, MeV$. In
particular the interference between the two wave functions in
eq.(\ref{gam}) significantly enhances the discussed decay if the $C$
parity of $\x$ is positive ($\eta=+1$) and suppresses the rate in the
case of negative $C$ parity ($\eta=-1$).

\begin{figure}[ht]
  \begin{center}
    \leavevmode
\epsfxsize=4in
\epsfbox{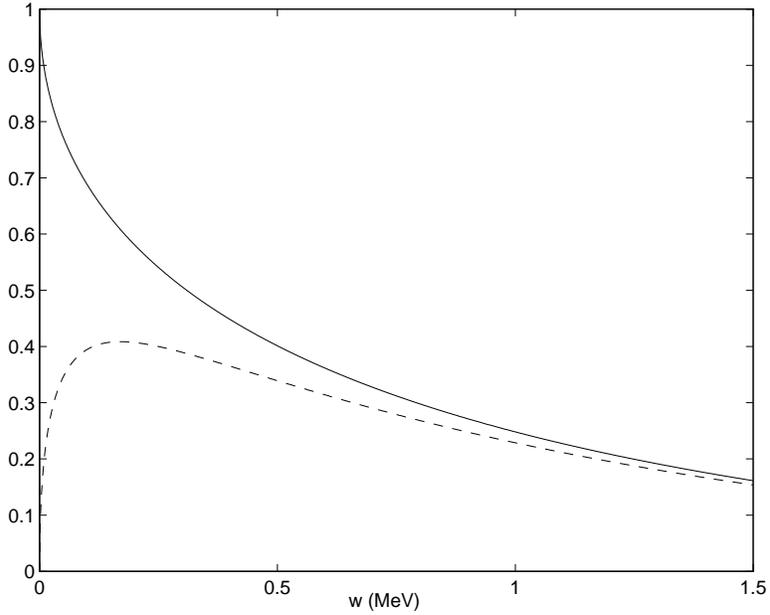}
\caption{The non-coherent contribution $A(w)$ (solid line) and the
interference term $B(w)$ (dashed) as defined in eq.(\ref{gamt}),
calculated by a numerical integration in eq.(\ref{gam}).}
\end{center}
\label{fig:xy}
\end{figure}

The sign of the interference term is reversed in the radiative decay $\x
\to D^0 {\bar D}^0 \gamma$ due to the negative $C$ parity of the photon.
The general expression for the decay rate, analogously to
eq.(\ref{gam}), has the form
\begin{eqnarray}
&&\!\!\!d \Gamma(X \to D^0 {\bar D}^0 \gamma)=  \\
&& \!\!\!{\Gamma_\gamma \over k_0^3} \, {k^2 \over  (2 \pi)^4} \, \left
| \phi({\vec q_2}) - \eta \, \phi({\vec q_1}) \right | ^2 \, \delta
\left (\Delta -
w - k - {q_1^2 \over 2 m }- {q_2^2 \over 2 m}  \right ) \, \delta^{(3)}
\left ( {\vec q_1}+{\vec q_2}+{\vec k} \right ) \, d^3   q_1 \, d^3 q_2
{d^3 k \over 2 k }~, \nonumber
\label{gamg}
\end{eqnarray}
where ${\vec k}$ is the momentum of the photon, $\Gamma_\gamma$ is the
width of the `free' decay $D^{*0} \to D^0 \gamma$ (from the available
data one can estimate $\Gamma_\gamma \approx 26 \pm 7 \, KeV$) , and
$k_0 \approx 137
\, MeV$ is the photon energy in the `free' decay. Since the binding
energy $w$ is in any case very small in comparison with $\Delta$, one
can neglect the small shift in the energy $k$ of the photon in the decay
of $\x$ in comparison with $k_0$. Furthermore, the effect of the recoil
of the heavy mesons, when their momentum changes on the scale of both
$\kappa$ and $k_0$, contributes very little to the energy balance, and
one can perform the integration over one of the heavy meson momenta by
neglecting the kinematical constraint on it. Also making use of the
normalization condition: $\int |\phi({\vec q})|^2 d^3 q/(2\pi)^3=
\xi^2$, one readily arrives at the following expression for the total
rate
\beq
\Gamma(X \to D^0 {\bar D}^0 \gamma)=2 \xi^2\, \Gamma_\gamma \, \left [1-
{\eta \over \xi^2} \, \int \phi({\vec k}+{\vec q}) \, \phi({\vec q}) \,
{d^3 q \over (2 \pi)^3 } \right ]~.
\label{gg}
\eeq
Using the expression (\ref{msw}) for the momentum space wave function,
one finally finds
\beq
\Gamma(X \to D^0 {\bar D}^0 \gamma)=2 \xi^2 \, \Gamma_\gamma \, \left (
1- \eta \, {2 \, \kappa \over k_0} \arctan {k_0 \over 2 \, \kappa}
\right )~.
\eeq
Clearly, the interference term described by this formula is quite
substantial even at very moderate values of $\kappa/k_0$: e.g. it
amounts to $0.32$ already at $w=0.1 \, MeV$, and to $0.71$ at $w=0.5 \,
MeV$.

One might argue that the integral for the interference term in
eq.(\ref{gg}) is mainly contributed by the behavior of the wave function
at momenta of order $q \sim k_0/2 \approx 70 MeV$, i.e. larger than
$\kappa$. However the corresponding distances $r \sim 2/k_0$ are still
beyond the range of strong forces, and the `free' approximation
(\ref{swave}) for the wave function should still be applicable.

As is already discussed, the free motion wave function (\ref{swave}) is
justified only at distances beyond the range of strong interactions, and
thus it fails to properly describe the dynamics at shorter distances, $r
\, \lsim \, m_\pi^{-1}$. At those distances, i.e. in the `core' of the
system, the mesons strongly overlap, and the whole `molecular' picture
of individual heavy mesons is likely to be inapplicable. It is not known
at present, how significant the non-molecular core part of the wave
function is, and in particular what is its contribution to the
amplitudes of the discussed decays. It is clear however that the
possible core contribution to these decays should lead to significantly
larger than $\kappa$ values of the momentum transfer to the final heavy
mesons in the decays. Thus this contribution can be revealed by studying
the momentum distribution of the $D$ and ${\bar D}$ mesons produced in
the decays. The core contribution should rather uniformly populate the
Dalitz plot, including the events, where both heavy mesons recoil with a
momentum significantly larger than $\kappa$, up to the kinematical
limits of the Dalitz plot. On the contrary, the discussed here
`molecular' contribution mainly populates the regions, where one of the
heavy mesons (the spectator) has a recoil momentum of order $\kappa$. In
other words, a study of the Dalitz plot of the discussed decays would
allow to literally scan the internal structure of the $\x$ resonance
and, possibly, to see both the molecular and core components of its
internal dynamics.

When this work was finished, there appeared the paper \cite{ps}, where
possible properties of the $\x$ resonance are discussed in connection
with a further study of the decay $\x \to \pi^+ \pi^- J/\psi$, including
the possibility of this resonance being dominantly a molecular type
state.

This work is supported in part
by the DOE grant DE-FG02-94ER40823.

\end{document}